# Theoretical foundation for the Pareto distribution of international trade strength and introduction of an equation for international trade forecasting

Mikrajuddin Abdullah

*Department of Physics, Bandung Institute of Technology, Jalan Ganesa 10, Bandung, 40132, Indonesia*

*Email: mikrajuddin@gmail.com*

ABSTRACT

I propose a new terminology, international trade strength, which is defined as the ratio of a country's total international trade to its GDP. This parameter represents a country's ability to generate international trade by utilizing its GDP. This figure is equivalent to GDP per capita, which represents a country's ability to use its population to generate GDP. Trade strength varies by country. The intriguing question is, what distribution function does the trade strength fulfill? In this paper, a theoretical foundation for predicting the distribution of trade strength and the rate of change of trade strength were developed. These two quantities were found to satisfy the Pareto distribution function. The equations were confirmed using data from the World Integrated Trade Solution (WITS) and the World Bank by comparing the Akaike Information Criterion (AIC) and Bayesian Information Criterion (BIC) to five types of distribution functions (exponential, lognormal, gamma, Pareto, and Weibull). I also discovered that the fitting Pareto power parameter is fairly close to the theoretical parameter. In addition, a formula for forecasting a country's total international trade in the following years was also developed.

*JEL classification*:
C10
C46
F10

## 1. Introduction

The law of gravity, in its modified form, has been successfully applied to explain international trade between countries in the last few decades (Anderson, 2011; Nijkamp and Ratajczak, 2021; Navarrete and Tatlonghari, 2018; Subhani et al, 2011; Kalirajan, 2007; Gul and Yasin, 2011). The equation was adopted from the Newton's law of gravity (Tinbergen, 1962; Linnemann, 1966), in which the volume of trade between countries is proportional to the multiplication of GDPs and inversely proportional to the square of the distance between two countries (distance to capital cities (Wall, 1999; Egger, 2000). However, trading volume is more complex to explain with the traditional gravity equation, so a number of changes have been suggested to ensure that the theory accurately explains the empirical evidence. Among the changes are adjustments to the power of GDP and the power of distance (Martin and Pham, 2020; Chaney, 2018; Wall, 1999; Gul and Yasin, 2011; Shahriar et al., 2019; Battersby and Ewing, 2005; Santos Silva and Tenreyro, 2006; Paas, 2000; Linders and de Groot, 2006; Matias, 2004; Haryadi and Hodijah, 2019; Kalirajan, 2007; Alleyne and Lorde, 2014; Effendi, 2014; Baldwin and Taglioni, 2011; Mnasri and Nechi, 2021), where the power of GDP is not always unity and the power of distance is not always two as in physics formula. Another proposed change is to include additional factors such as the country's population, GDP per capita, language, border effect (shared or non-shared border), and remoteness (Navarrete and Tatlonghari, 2018; Subhani et al, 2011; Kalirajan, 2007; Nijkamp and Ratajczak, 2021; Gul and Yasin, 2011). Finally, the resulting equation is difficult to recognize as the gravitational equation. An intriguing question is how to return the international trade equation to a nearly pure gravitational form, as introduced by Tinbergen (1962). If we pursue accuracy by following current trends, the derived equations will continue to deviate from the pure gravity equation due to the introduction of new components.

We previously proposed that the Coulomb theory more precisely explains trade volume between countries than the gravity theory (Abdullah, 2023). Until now, no researcher has used the Coulomb force equation to explain international trade. The coulomb force between charges has an equation that is very similar to the gravitational force between masses. The only difference between the two forces is that the coulomb force has a material effect, which is expressed as the dielectric constant. Gravitational forces do not have a similar effect. Even if the product of the two charges and the square of the distance is equal, the Coulomb force can differ if the dielectric constants differ.

In order to return the modified gravitational equation to its simple form, we can replace this equation with the Coulomb equation in which all the variables affecting trading volume, except charge and distance, are grouped under one dielectric constant. The dielectric constant takes into account factors like population size, GDP per capita, border effects, political relations between the two countries, economic agreements between the two countries, and so on (Abdullah, 2023). The dielectric constant can vary between pairs of countries and change over time.

In gravitational or coulomb forces, we defined the variable strength of the force at a point. For the gravitational force, the acceleration due to gravity represents the gravitational force's strength. Only external factors influence the strength of a force at a point, which determines the gravitational force at that point if a mass is placed there. A comparable variable should be definable in the international trade. A country's trading strength can be defined as the sum of the effects that all other countries have on that country's position. When this strength is multiplied by the country's GDP, the total trade of the country is produced. The trading strength of a country is determined not by its own economy, but by the influence of all other countries. Certain countries only use their strength to generate total trade, and they can control total trade only by controlling their GDP.

Because each country's GDP is different, so is each country's trading strength. As a result, it is worthwhile to consider how the world's trading strength is distributed among countries. Furthermore, because each country's GDP is time-dependent in general, trading strength in specific positions will be time-dependent as well. Finally, we have the rate of change in each country's trading strength. The next question is, how is the rate of trading strength distributed? To the best of our knowledge, researchers have not reported on studies of the distribution of these quantities. Both of these quantities can be used to forecast the growth of a country's international trade or to determine the best ways to increase each country's trade.

The purpose of this paper is to derive the distribution functions and rates of trading strength of countries. We employ Chandrasekhar and Neumann's (CN) (Chandrasekhar and von Neumann, 1942) explanation of gravitational fluctuations. In particular, we begin with the coulomb equation (Abdullah, 2023) because its form is much simpler than the modified gravity equation. We also used machine learning to group countries into clusters based on GDP and trade strength.

## 2. Method

We have shown that the coulomb equation accurately describes international trade better than the gravitational equation (Abdullah, 2023). We also estimated that the power for GDP and distance is approximately one, as previously claimed by Chaney (2018), yielding to the equation

$$F_{mn}(t) = K \frac{G_m(t) G_n(t)}{\kappa_{mn}(t) R_{mn}} \quad (1)$$

where $G_m(t)$ denotes GDP in year $t$, $K$ is a constant, and $\kappa_{mn}(t)$ denotes the "dielectric constant" in year $t$ between countries $m$ and $n$. As previously discussed (Abdullah, 2023), $F_{mn}$ is a symmetric variable. We definie the "trade strength" in the position of country $m$ generated by country $n$'s GDP as

$$g_{mn}(t) = \frac{F_{mn}(t)}{G_m(t)} = K \frac{G_n(t)}{\kappa_{mn}(t) R_{mn}} \quad (2)$$

Equation (2) is similar to the "income potential" equation proposed by W. Isard (Capoani, 2023). The "trade strength" due to the country with GDP $G(t)$ at position $R$ in the continuum variable is given by

$$g(t) = K \frac{G(t)}{\kappa(t) R} \quad (3)$$

The $g$ variable simulates the acceleration caused by gravity.

The GDP of a country always changes over time (World Bank,2023). The general trend is that GDP rises over time, although it may fall in some years due to the emergence of global economic problems. For example, Japan's GDP fell between 1995 and 1998 and again between 2012 and 2015 (World Bank,2023). Let us write

$$G(t) = G(t_0)\phi(t_0, t) \tag{4}$$

where $G(t_0)$ is GDP in year $t_0$, and $\phi(t_0, t)$ is a time-dependent function that satisfies $\phi(t_0, t_0) = 1$. By substituting Eq. (4) into Eq. (3), one gets

$$g(t) = \frac{G(t_0)}{\kappa(t)R/K\phi(t_0,t)}$$

$$= \frac{G(t_0)}{T(t)} \tag{5}$$

where

$$T(t) = \frac{\kappa(t)R}{K\phi(t_0,t)} \tag{6}$$

We can call $T$ the "diplomatic distance" between countries, and it is affected by the spatial distance, dielectric constant, and changes in GDP. The rate of change of $g(t)$ with respect to time is

$$f(t) = \frac{dg}{dt} = -\frac{G(t_0)}{T^2}\frac{dT}{dt}$$

$$= -\frac{G(t_0)}{T^2}v \tag{7}$$

where $v = dT/dt$.

We write the probability of trade strength at distances $T$ to $T + dT$ and speeds $v$ to $v + dv$ as $\tau(T, v)dTdv$. The probability distribution of $g$ in the range of $g_0 \leq g \leq g_0 + dg_0$ and $f$ in the range of $f_0 \leq f \leq f_0 + df_0$ is

$$w(g_0, f_0, G)dg_0 df_0 = \frac{1}{A}\int\int_{g_0 \leq g \leq g_0+dg_0, f_0 \leq f \leq f_0+df_0} \tau(T,v)dTdv \tag{8}$$

where $A$ represents the total "phase" area of all countries. Let us use the Dirichlet factor, which has a value of one if the preceding constraint is met and zero otherwise, i.e. (Chandrasekhar and von Neumann, 1942, von Laue, 1915)

$$\frac{1}{\pi^2}\int_{-\infty}^{\infty}\int_{-\infty}^{\infty}e^{i\rho(g-g_0)}e^{i\sigma(f-f_0)}\frac{\sin\left(\frac{1}{2}\rho g\right)}{\rho}\frac{\sin\left(\frac{1}{2}\sigma f\right)}{\sigma}d\rho d\sigma = \begin{cases} 1 & \text{if} \quad g_0 \leq g \leq g_0 + +dg_0 \\ & \quad\quad\text{and } f_0 \leq f \leq f_0 + df_0 \\ 0 & \text{if} \quad\quad\quad\quad others \end{cases}$$

(9)

Using Eq. (9), we can write Eq. (8) as

$$w(g_0, f_0, G)dg_0 df_0 = \frac{dg_0 df_0}{4\pi^2}\int_{-\infty}^{\infty}\int_{-\infty}^{\infty}e^{-i\rho g_0}e^{-i\sigma f_0}B(\rho, \sigma, G)d\rho d\sigma \quad (10)$$

where

$$B(\rho, \sigma, G) = \frac{1}{A}\int_{T=T_{min}}^{T_{max}}\int_{v=-\infty}^{\infty}e^{i\rho g}e^{i\sigma f}\tau(T, v)dTdv$$

$$\approx \frac{1}{A}\int_{0}^{\infty}\int_{-\infty}^{\infty}e^{i\rho g}e^{i\sigma f}\tau(T, v)dTdv \quad (11)$$

To explain the fluctuations in gravitational strength, CN (1942) assumes that the probability is solely determined by speed using the equation $\tau = qe^{-p^2v^2}$, where $p$ and $q$ are constants. We employ a more general equation in which the diplomatic distance influences the probability. Because the probability should decrease with increasing diplomatic distance, we propose a general form

$$\tau = \frac{q}{T^\theta}e^{-p^2v^2} \quad (12)$$

where $p$, $q$, and $\theta$ are all positive constants. When we substitute Eq. (12) into Eq. (11), we get

$$B(\rho, \sigma, G) = \frac{q}{A}\int_{0}^{\infty}\frac{1}{T^\theta}e^{i\rho g}dT\int_{-\infty}^{\infty}\int e^{-i\sigma\frac{G}{T^2}v}e^{-p^2v^2}dv \quad (13)$$

Let us use identity (Chandrasekhar, S. and von Neumann, 1942)

$$\int_{-\infty}^{\infty}e^{-iQ\omega - (1/4\varphi^3)\omega^2}d\omega = 2\sqrt{\pi}\varphi^{3/2}e^{-Q^2\varphi^3} \quad (14)$$

so we get

$$B(\rho, \sigma, G) = \frac{q}{A}\frac{\sqrt{\pi}}{p}\int_{0}^{\infty}\frac{1}{T^\theta}e^{i\rho\frac{G}{T}}\exp\left[-\frac{G^2\sigma^2}{4p^2T^4}\right]dT \quad (15)$$

To solve Eq. (15), let us assume

$$x^4 = \frac{G^2\sigma^2}{4p^2T^4} \tag{16}$$

Using Eqs. (15) and (16) and a series of mathematical steps, we obtain

$$B(\rho,\sigma,G) = \frac{q\sqrt{\pi}}{Ap}\left(\frac{2p}{G}\right)^{\frac{\theta-1}{2}} \sigma^{\frac{1-\theta}{2}} \int_0^\infty x^{\theta-2} e^{-i\rho\sqrt{2pG/\sigma}x} e^{-x^4} dx \tag{17}$$

Then, we substitute Eq. (17) into Eq. (10), replacing the variables $g_0 \to g$ and $f_0 \to f$ in Eq. (10) (Chandrasekhar and von Neumann, 1942), one gets

$$w(g,f,G) = \frac{1}{4\pi^2} \int_{-\infty}^\infty \int_{-\infty}^\infty e^{-i\rho g} e^{-i\sigma f} \left\{ \frac{q\sqrt{\pi}}{Ap}\left(\frac{2p}{G}\right)^{\frac{\theta-1}{2}} \sigma^{\frac{1-\theta}{2}} \int_0^\infty x^{\theta-2} e^{i\rho\sqrt{2pG/\sigma}x} e^{-x^4} dx \right\} d\rho d\sigma$$

$$= \frac{1}{4\pi^2} \frac{q\sqrt{\pi}}{Ap}\left(\frac{2p}{G}\right)^{\frac{\theta-1}{2}} \int_{-\infty}^\infty e^{-i\sigma f} \sigma^{\frac{1-\theta}{2}} \int_0^\infty x^{\theta-2} e^{-x^4} \left( \int_{-\infty}^\infty e^{-i\rho(g-\sqrt{2pG/\sigma}x)} d\rho \right) dx\, d\sigma \tag{18}$$

We integrate the variable $\rho$ and use the Dirac delta definition

$$\delta(z-z_0) = \frac{1}{2\pi}\int_{-\infty}^\infty \exp(-i(z-z_0)\omega)\, d\omega \tag{19}$$

to get

$$w(g,f,G) = \frac{1}{4\pi^2} \frac{q\sqrt{\pi}}{Ap}\left(\frac{2p}{G}\right)^{\frac{\theta-1}{2}} \int_{-\infty}^\infty e^{-i\sigma f} \sigma^{\frac{1-\theta}{2}} \int_0^\infty x^{\theta-2} e^{-x^4} \left( 2\pi\delta\left(g-\sqrt{2pG/\sigma}x\right) \right) dx\, d\sigma \tag{20}$$

Furthermore, we use Dirac's delta property $\delta(\alpha(z-z_0)) = (1/|\alpha|)\delta((z-z_0))$ to obtain

$$w(g,f,G) = \frac{1}{4\pi^2} \frac{q\sqrt{\pi}}{Ap}\left(\frac{2p}{G}\right)^{\frac{\theta-1}{2}} \int_{-\infty}^\infty e^{-i\sigma f} \sigma^{\frac{1-\theta}{2}} \int_0^\infty x^{\theta-2} e^{-x^4} \left( \frac{2\pi}{\sqrt{2pG/\sigma}}\delta\left(x - \frac{g}{\sqrt{2pG/\sigma}}\right) \right) dx\, d\sigma$$

$$= \frac{1}{2\pi} \frac{q\sqrt{\pi}}{Ap\sqrt{2pG}}\left(\frac{2p}{G}\right)^{\frac{\theta-1}{2}} \int_{-\infty}^\infty e^{-i\sigma f} \sigma^{1-\frac{\theta}{2}} \int_0^\infty x^{\theta-2} e^{-x^4} \delta\left(x - \frac{g}{\sqrt{2pG/\sigma}}\right) dx\, d\sigma \tag{21}$$

When integrating over $x$, one obtains

$$w(g,f,G) = \frac{1}{2\pi} \frac{q\sqrt{\pi}}{Ap\sqrt{2pG}}\left(\frac{2p}{G}\right)^{\frac{\theta-1}{2}} \int_{-\infty}^\infty e^{-i\sigma f} \sigma^{1-\frac{\theta}{2}} \left(\frac{g}{\sqrt{2pG/\sigma}}\right)^{\theta-2} \exp\left[-\left(\frac{g}{\sqrt{2pG/\sigma}}\right)^4\right] d\sigma$$

$$= \frac{1}{2\pi} \frac{q\sqrt{\pi}}{Ap\sqrt{2pG}}\left(\frac{2p}{G}\right)^{\frac{\theta-1}{2}} \left(\frac{g}{\sqrt{2pG}}\right)^{\theta-2} \int_{-\infty}^\infty e^{-i\sigma f - V\sigma^2} d\sigma \tag{22}$$

where $V = g^4/4p^2 G^2$. Again, using the identity in Eq. (14), one obtains

$$w(g,f,G) = \Omega \frac{G^{2-\theta}}{g^{4-\theta}} \exp\left(-\frac{p^2 G^2 f^2}{g^4}\right) \qquad (23)$$

We now examine how the variations of $w$ with respect to $g$ and $G$ alone and with respect to $f$ ang $G$ alone, performed by integrating $f$ (from $-\infty$ to $+\infty$) and $g$ (from 0 to $+\infty$). We obtain

$$w(g,G) = \int_{-\infty}^{\infty} w(g,f,G) df \propto \frac{G^{1-\theta}}{g^{2-\theta}} \qquad (24)$$

$$w(f,G) = \int_0^{\infty} w(g,f,G) dg \propto \frac{G^{1/2-\theta/2}}{|f|^{3/2-\theta/2}} \qquad (25)$$

Equation (24) expresses the $g$ distribution produced by a given $G$ at any position between $T = 0$ and $T \to \infty$. Similarly, Eq. (25) expresses the $f$ distribution produced by a given $G$ at any position between $T = 0$ and $T \to \infty$. The resulting distribution of all possible values of $G$ can be obtained by averaging Eqs. (24) and (25) over $G$ as

$$w(g) = \overline{W(g,G)} \propto \overline{\frac{G^{1-\theta}}{g^{2-\theta}}} \qquad (26)$$

$$w(f) = \overline{w(f,G)} \propto \overline{\frac{G^{1/2-\theta/2}}{|f|^{3/2-\theta/2}}} \qquad (27)$$

Equations (26) and (27) clearly resemble the Pareto distribution function (Warsono et al, 2019; Bourguignon et al, 2022; Zaninetti and Ferraro, 2008), which has the general form,

$$h(x) = \begin{cases} 0 & \text{if } x < \beta \\ \frac{\alpha \beta^\alpha}{x^{\alpha+1}} & \text{if } x \geq \beta \end{cases} \qquad (28)$$

where $\beta > 0$. Equation (26) corresponds to $\alpha = 1 - \theta$ and Eq. (27) corresponds to $\alpha = (1 - \theta)/2$. In the Results and Discussion section, we will show that the data from WITS (2023) and World Bank (2023) satisfy the Pareto distribution function by calculating the AIC (Akaike, 1973) and the BIC (Claeskens and Hjort, 2008). We compute the AIC and BIC of various distribution functions (exponential, lognormal, gamma, Pareto, and Weibull) and examine which distribution has the smallest AIC and BIC values.

## 3. Results and Discussion

### 3.1 Validating the Pareto distribution function

We will validate Eqs. (24) and 25) using real-world data. We use WITS (2023) trade data from 2004, 2009, 2014, and 2019. We use World Bank (2023) data for GDP and GDP growth for 2004, 2009, 2014, and 2019. Trade strength is calculated by the equation

$$g_m = \sum_n g_{mn} = \frac{\sum_n F_{mn}}{G_m} \qquad (29)$$

(see Eq. (2)). According to the equation above, $\sum_n F_{mn}$ represents world trade for country $m$.

We obtained trade data for all countries from the WITS website in 2004, 2009, 2014, and 2019 (WITS, 2023). Every year, approximately 250 files are downloaded. Each country's data lists other countries that trade with it in the form of exports, imports, trade balances, and so on. There is also aggregate data (World trade) that we will use as $\sum_n F_{mn}$, the sum of global exports and imports of each country. The Python program is used to extract data from all files.

For a given year, the value of $g$ ranges from very small to very large. Furthermore, it is intriguing to investigate which type of distribution function the $g$ variable fulfills. Because $g$ is always positive, possible distribution functions include exponential, lognormal, gamma, Pareto, and Weibull. To determine the best fit distribution function, we compute the AIC and BIC from the equations

$$AIC = 2k - 2LL_{max} \qquad (30)$$

$$BIC = k \ln n - 2LL_{max} \qquad (31)$$

where $k$ is the number of distribution function parameters, $n$ is the number of data points, and $LL_{max}$ is the natural logarithm of the maximum likelihood (see **Appendix 1**). The distribution function to be used is the one with the smallest AIC or BIC.

Table 1 displays AIC and BIC calculations for data from 2004, 2009, 2014, and 2019. We conclude that the $g$ distribution better satisfies the Pareto distribution because it appears to have the smallest AIC and BIC. We also discovered that the Pareto parameter ranges from 0.66 to 0.82.

**Table 1**

A comparison of the results of AIC and BIC calculations for various distribution functions $W(g)$

| Distribution Function | Information Criterion | Year | | | |
|---|---|---|---|---|---|
| | | 2004 | 2009 | 2014 | 2019 |
| Exponential | AIC | -1989.70 | -1950.40 | -1970.22 | -1736.81 |
| | BIC | -1981.55 | -1942.30 | -1962.07 | -1733.39 |
| Gamma | AIC | -2024.15 | -1984.85 | -2005.20 | -1819.63 |
| | BIC | -2018.00 | -1978.75 | -1999.05 | -1813.59 |
| Lognormal | AIC | -2102.02 | -2072.47 | -2090.53 | -1832.42 |
| | BIC | -2095.87 | -2066.37 | -2084.38 | -1826.39 |
| **Pareto** | AIC | **-2295.28** | **-2194.00** | **-2234.44** | **-1992.82** |
| | BIC | **-2289.13** | **-2187.90** | **-2228.29** | **-1986.79** |
| | $\alpha$ | 0.82 | 0.66 | 0.70 | 0.69 |
| | $\beta$ | $1.86 \times 10^{-4}$ | $1.37 \times 10^{-4}$ | $1.62 \times 10^{-4}$ | $2.31 \times 10^{-4}$ |
| Weibull | AIC | -2070.46 | -2042.43 | -2056.50 | -1795.76 |
| | BIC | -2064.30 | -2036.33 | -2050.35 | -1789.73 |

To determine the distribution of $f$, we must first estimate the value of $f$. To accomplish this, we return to the initial equation, which can be written as

$$g(t) = K \frac{G(t)}{\kappa(t)R} \tag{32}$$

We do not express $G$ in terms of the Eq. (4). If we assume that $\kappa(t)$ changes very slowly with time under normal conditions, we get from Eq. (32) approximately

$$f = \frac{dg}{dt} \approx K \frac{KG(t)}{\kappa(t)R} \left( \frac{1}{G(t)} \frac{dG}{dt} \right)$$

$$= g G_{gr} \tag{33}$$

where $G_{gr}$ is growth rate of the GDP.

Table 2 displays the results of calculating the AIC and BIC distributions $f$ for the five distribution functions. We also discover that the Pareto distribution function produces the smallest AIC and BIC values, implying that the Pareto distribution function more accurately describes the $f$ distribution. We also discovered that the Pareto parameter $\alpha$ falls between 0.16 and 0.33.

The value of $\alpha$ in Tables 1 and 2 can be used to estimate the parameter in Eqs. (26) and (27). If we choose $\theta \approx 0.5$, the distribution of $g$ changes by $\propto 1/g^{1.5}$ and the distribution of $f$ changes by $\propto 1/|f|^{1.25}$. This means that we get $\alpha \approx 0.5$ for the $g$ distribution and $\alpha \approx 0.25$ for the $f$ distribution. The two results above are surprising. The powers obtained theoretically are comparable to the powers obtained from real data.

**Table 2**

Comparison of AIC and BIC calculation results for various $W(f)$ distribution functions

| Distribution Function | Information Criterion | Year | | | |
|---|---|---|---|---|---|
| | | 2004 | 2009 | 2014 | 2019 |
| Exponential | AIC | -2864.82 | -2889.53 | -2888.74 | -2724.36 |
| | BIC | -2861.76 | -2881.46 | -2.880.72 | -2721.34 |
| Gamma | AIC | -2870.46 | -2888.68 | -2712.26 | -2722.59 |
| | BIC | -2864.34 | -2882.61 | -2706.24 | -2716.56 |
| Lognormal | AIC | -2867.94 | -2878.43 | -2941.58 | -2693.88 |
| | BIC | -2861.81 | -2872.36 | -2935.56 | -2687.96 |
| **Pareto** | **AIC** | **-3095.45** | **-3113.14** | **-3233.79** | **-2959.13** |
| | **BIC** | **-3089.32** | **-3107.07** | **-3227.37** | **-2953.11** |
| | $\alpha$ | 0.33 | 0.22 | 0.16 | 0.18 |
| | $\beta$ | $1.29 \times 10^{-6}$ | $2.04 \times 10^{-7}$ | $1.99 \times 10^{-8}$ | $8.74 \times 10^{-8}$ |
| Weibull | AIC | -2867.57 | -2887.88 | -2800.16 | -2722.67 |
| | BIC | -2861.44 | -2881.80 | -2794.14 | -2716.65 |

Another interesting finding is that the distribution of GDP also fulfills the Pareto function with $\alpha \approx 0.135$. Shown in **Appendix 2** is the comparison of the results of AIC and BIC calculations for GDP in 2005, 2010, 2015, and 2020. Similar calculations have been reported by Tomić (2018). The parameter $\beta$ takes $G_{min}$ (the minimum value of GDP) so the average values in Eqs. (26) and (27) become

$$\overline{G^{1-\theta}} = \frac{\int_{G_{min}}^{G_{max}} G^{1-\theta} \frac{\alpha G_{min}^{\alpha}}{G^{1+\alpha}} dG}{\int_{G_{min}}^{G_{max}} \frac{\alpha G_{min}^{\alpha}}{G^{1+\alpha}} dG} = \left(\frac{\alpha}{1-\theta-\alpha}\right) \frac{G_{max}^{1-\theta-\alpha} - G_{min}^{1-\theta-\alpha}}{G_{min}^{-\alpha} - G_{max}^{-\alpha}} \qquad (34)$$

$$\overline{G^{1/2-\theta/2}} = \frac{\int_{G_{min}}^{G_{max}} G^{1/2-\theta/2} \frac{\alpha G_{min}^{\alpha}}{G^{1+\alpha}} dG}{\int_{G_{min}}^{G_{max}} \frac{\alpha G_{min}^{\alpha}}{G^{1+\alpha}} dG} = \left(\frac{\alpha}{1/2-\theta/2-\alpha}\right) \frac{G_{max}^{1/2-\theta/2-\alpha} - G_{min}^{1/2-\theta/2-\alpha}}{G_{min}^{-\alpha} - G_{max}^{-\alpha}} \qquad (35)$$

Using $\theta \approx 0.5$ and $\alpha \approx 0.135$ and by approximating $G_{max} \gg G_{min}$, we get $\overline{G^{1-\theta}} \propto G_{max}^{0.365} G_{min}^{0.135}$ and $\overline{G^{1/2-\theta/2}} \propto G_{max}^{0.115} G_{min}^{0.135}$, so Eqs. (26) and (27) become

$$w(g) \propto \frac{G_{max}^{0.365} G_{min}^{0.135}}{g^{3/2}} \qquad (36)$$

$$w(f) \propto \frac{G_{max}^{0.115} G_{min}^{0.135}}{|f|^{5/4}} \qquad (37)$$

Figure 1 shows a comparison of the CDF for $g$ and the Pareto distribution function calculation results using $\alpha = 0.5$. Because $CDF = 1$ when $g = g_{max}$ and the Pareto function equals one only when $g \to \infty$, we scale the Pareto's CDF so that it is $\approx 1$ when $g = g_{max}$. Figure 1 shows that the curve does not exactly coincide with the data, but it is good enough to explain the variation in the data. According to the results of our AIC and BIC calculations, the Pareto distribution function should better explain data variation than other distribution functions.

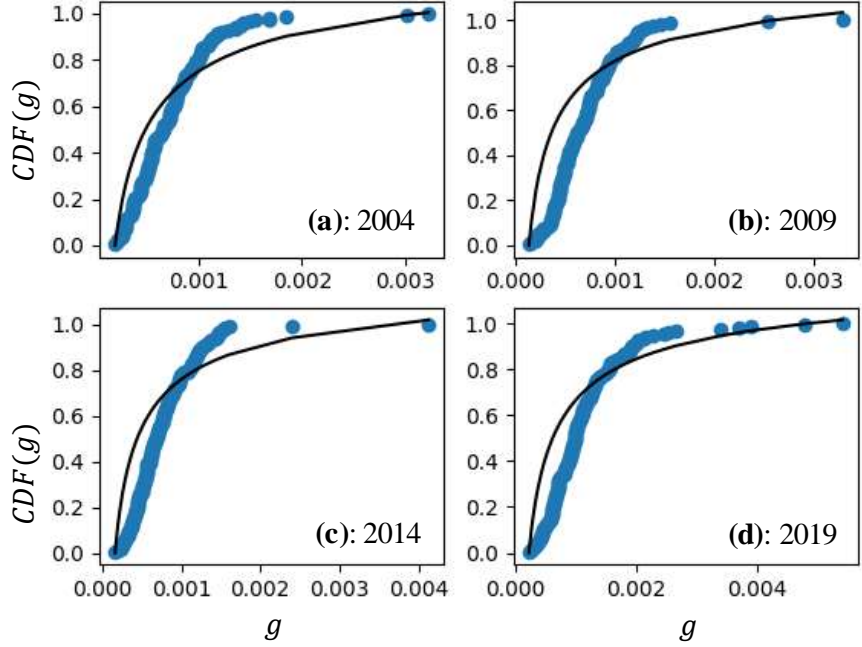

**Fig. 1** CDF from data (symbols) versus CDF calculated using the Pareto distribution function with $\alpha = 0.5$ (curve) in different years: (a) 2004, (b) 2009, (c) 2014, and (d) 2019.

### 3.2 Forecasting future total trade

We can get an estimate from Eq. (33)

$$\frac{dg}{dt} \approx g G_{gr} \tag{38}$$

Furthermore, we may deduce from Eq. (38) the dependence of $g$ on the time satisfies the equation

$$g(t) = g(t_0) \exp\left(\int_{t_0}^{t} G_{gr} dt\right) \tag{39}$$

Thus, the time dependence of a country's total trade satisfies the equation

$$F(t) = g(t)G(t) \approx g(t_0)G(t) \exp\left(\int_{t_0}^{t} G_{gr} dt\right) \tag{40}$$

Equation (40) states that, we can forecast a country's total international trade if we can predict the GDP and the GDP growth rate. These two variables can generally be forecasted for the next few years (Statista 2023), allowing a country's total international trade during the next few years to be predicted.

Before we go any farther, we need to see if Eq. (40) fits the empirical datas. We will compare total international trade data for numerous nations with the estimation findings from Eq. (40). To make the comparison results more representative, we will take a sample of eight countries, two each with GDP of the order of $10^9$ USD (Andorra and Suriname), GDP of the order of $10^{10}$ USD (Benin and Guatemala), GDP of the order of $10^{11}$ USD (Argentina and the Netherlands), and GDP of the order of $10^{12}$ USD (Japan and the United Kingdom), where such GDPs are in 2020 (World Bank, 2023). We used data spanning the years 1990 to 2020. Some countries, however, only have GDP data from 1992, 1993, and 1994. We replace the integral in Eq. (40) with the following summation

$$\int_{t_0}^{t} G_{gr} dt = \sum_{i=initial\ year}^{i=target\ year} G_{gr}(i) \times 1 year \tag{41}$$

Furthermore, we normalize the trading value into variable

$$\tilde{F}(i) = \frac{F(i) - F_{av}}{\sigma} \tag{42}$$

where $F_{av}$ represents the trading average and $\sigma$ is the standard deviation.

Figure 2 depicts a comparison of trade data from WITS (2023) (blue curve) and calculation results using Eq. (40) (orange curve) for a total of eight countries. We discovered a reasonably excellent agreement between the data and the estimated outcomes using Eq. (40). This demonstrates how Eq. (41) can be used to forecast future international trades. We can forecast a country's international trade value in the future if we can estimate its GDP and GDP growth rate.

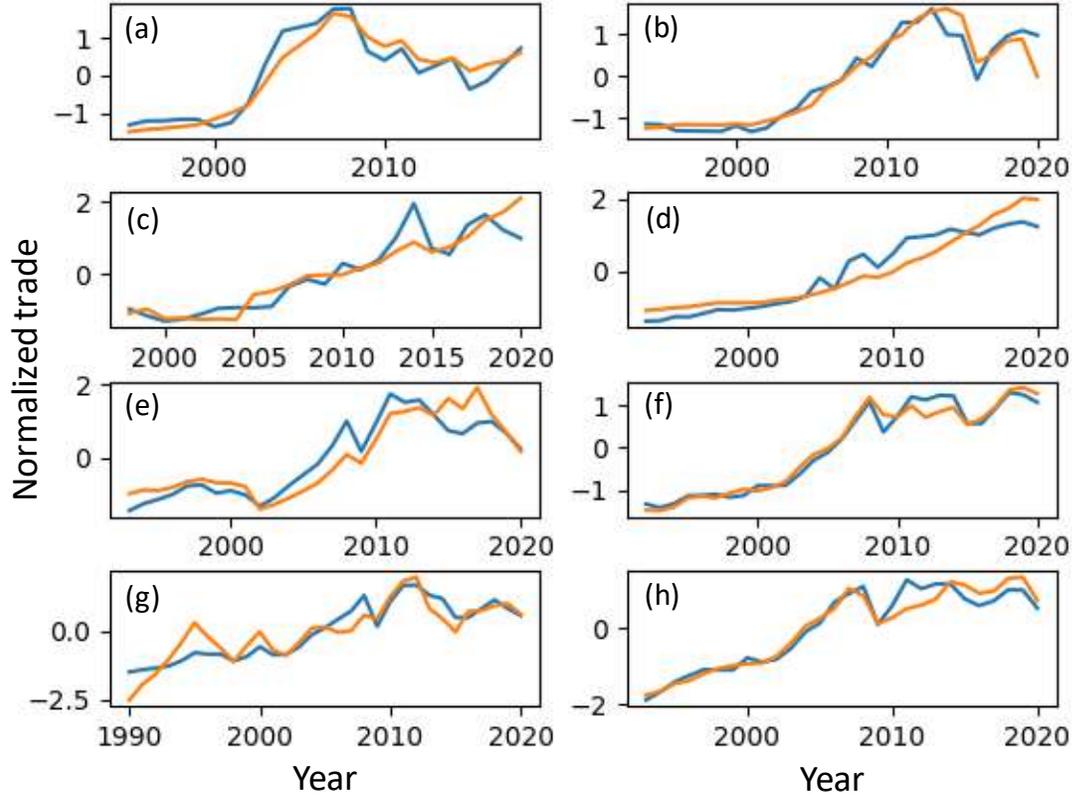

**Fig. 2** Comparison of WITS international trade statistics for a total of eight nations (blue curve) and calculation results using Eq. (40): (a) Andorra, (b) Suriname, (c) Benin, (d) Guatemala, (e) Argentina, (d) The Netherlands, (g) Japan, and (h) The United Kingdom. (blue curve) The data from WITS (2023) and (orange curve) are the results of estimation using Eq. (40). We show the normalizing variables given by Eq. (42).

If we take $t = t_0$, Eq. (40) becomes

$$F(t_0) \approx g(t_0)G(t_0) \exp\left(\int_{t_0}^{t_0} G_{gr} dt\right) = g(t_0)G(t_0) \quad (43)$$

Finally, Eq. (40) can be expressed in another way, namely

$$F(t) \approx F(t_0) \frac{G(t)}{G(t_0)} \exp\left(\int_{t_0}^{t} G_{gr} dt\right) \quad (44)$$

Equation (44) can be read as follows. Explicitly, a country's total international trade can be simply computed as long as we can estimate the change in the country's GDP over time and the growth rate of the country's GDP as a function of time. We no longer need to know other countries' GDPs and distances. Only when we wish to determine the trade value of a pair of countries do we need to know other country's GDP and distance.

Figure 3 depicts an estimate of international trade for eight countries. Trade values have been normalized to trade values in 2020. We estimate up to 2028 using Statista's GDP and GDP growth forecast (Statista, 2023). Each figure depicts two countries with comparable GDPs. Togo and Suriname in Figure (a) have 2020 GDP of about $10^9$ USD, Crotia and Guatemala in Figure (b) have 2020 GDP of around $10^{10}$ USD, Argentina and the Netherlands in Figure (c) have GDP of around $10^{11}$ USD, and Japan and the United Kingdom in Figure (d) have GDP of around $10^{12}$ USD (World Bank, 2023). According to these figures, only Japan shows a fall in trade, which occurred in 2022.

The question is whether these forecasts can be accepted. Take the United Kingdom as an example. The predicted total trade value for the UK in 2028 is almost 1.9 times that of trade in 2020 (WITS, 2023). The UK's GDP in 2020 is USD 2706.54 billion (World Bank, 2023; Statista, 2023), while a projected GDP of USD 4245.42 billion (Statista, 2023) for 2028. So, throughout the course of eight years, GDP increased by almost 1.59 times. Let us examine a comparable timeframe in which the GDP rose at a similar rate to 2020 to 2028, which occurred between 1999 and 2007. The UK's GDP in 1999 was USD 1.69 trillion, and it was USD 3.09 trillion in 2007 (World Bank, 2023), representing a 1.83-fold growth in GDP. The UK's international trade value was USD 581 billion in 1999 and USD 1.13 trillion in 2007 (WITS, 2023), representing a 1.95-fold growth. There appear to be commonalities in the growth of GDP and trade in these two timeframes.

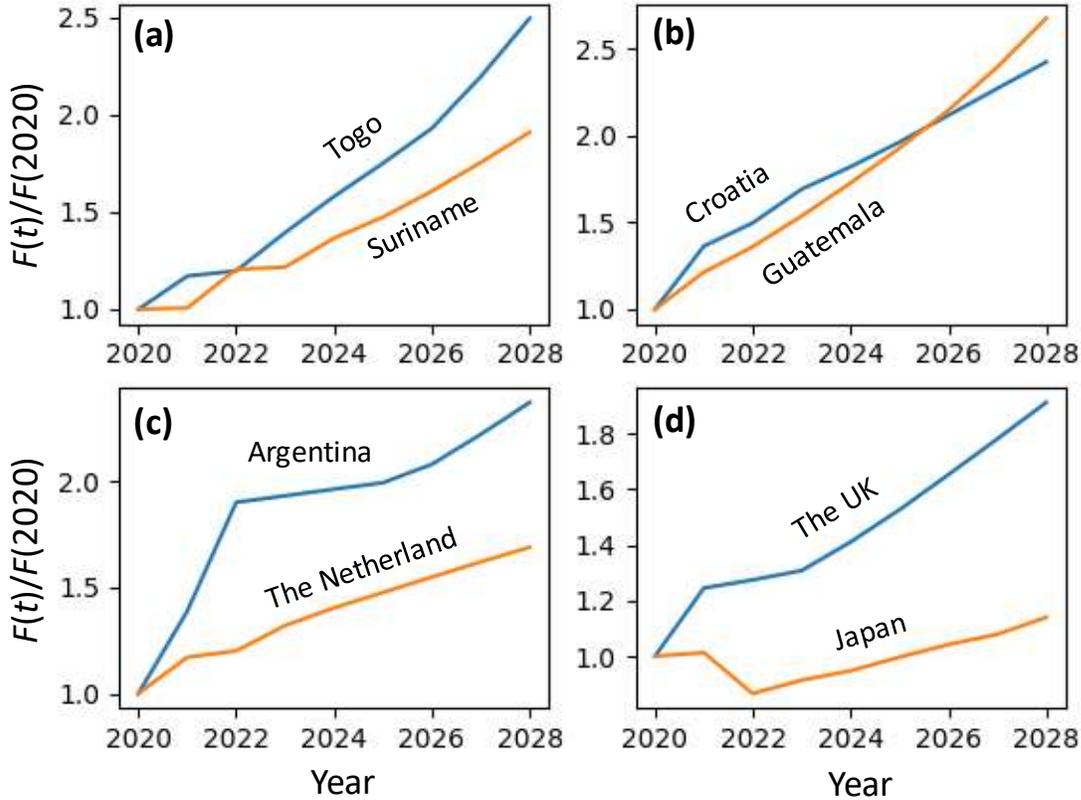

**Fig. 3** Estimated total international trade of the eight sample countries untill 2028. Figure (a) represents countries with 2020 GDPs of around $10^9$ USD, (b) depicts countries with 2020 GDPs of around $10^{10}$ USD, (c) depicts countries with 2020 GDPs of around $10^{11}$ USD, and (d) depicts countries with 2020 GDPs of around $10^{12}$ USD.

**3.3 Machine Learning Clustering**

Let us go over the definition of trade strength, $g$, in greater detail. Based on Eq. (29), we can deduce that $g$ represents a country's ability to use its GDP to generate international trade. This figure is equivalent to GDP per capita, which represents a country's ability to use its population to generate GDP. The higher the value of $g$, the better the country uses GDP to generate international trade. Intuitively, increasing GDP should increase the value of $g$, but this is not the case.

We will use unsupervised machine learning to cluster all countries based on value pairs $(G, g)$. For clustering data points, we will employ the *K-means* method, which is an unsupervised learning method. First, the data $g$ and $G$ are normalized using the following equations

$$\tilde{g}(i) = \frac{g(i) - g_{av}}{\sigma_g} \tag{45}$$

and

$$\tilde{G}(i) = \frac{G(i) - G_{av}}{\sigma_G} \tag{46}$$

where $g_{av}$ and $G_{av}$ are the averages for $g$ and $G$, respectively, and $\sigma_g$ and $\sigma_G$ are the standard deviations for $g$ and $G$, respectively.

The first step is to determine the number of appropriate clusters. For this purpose, we use the elbow method to calculate the inertia as a function of the number of clusters. Figure 4 depicts inertia as a function of the number of clusters for data $(G, g)$ from 2004 to 2019. The number of clusters of four appears to be the optimal number because inertia does not change significantly if the number of clusters is greater than four.

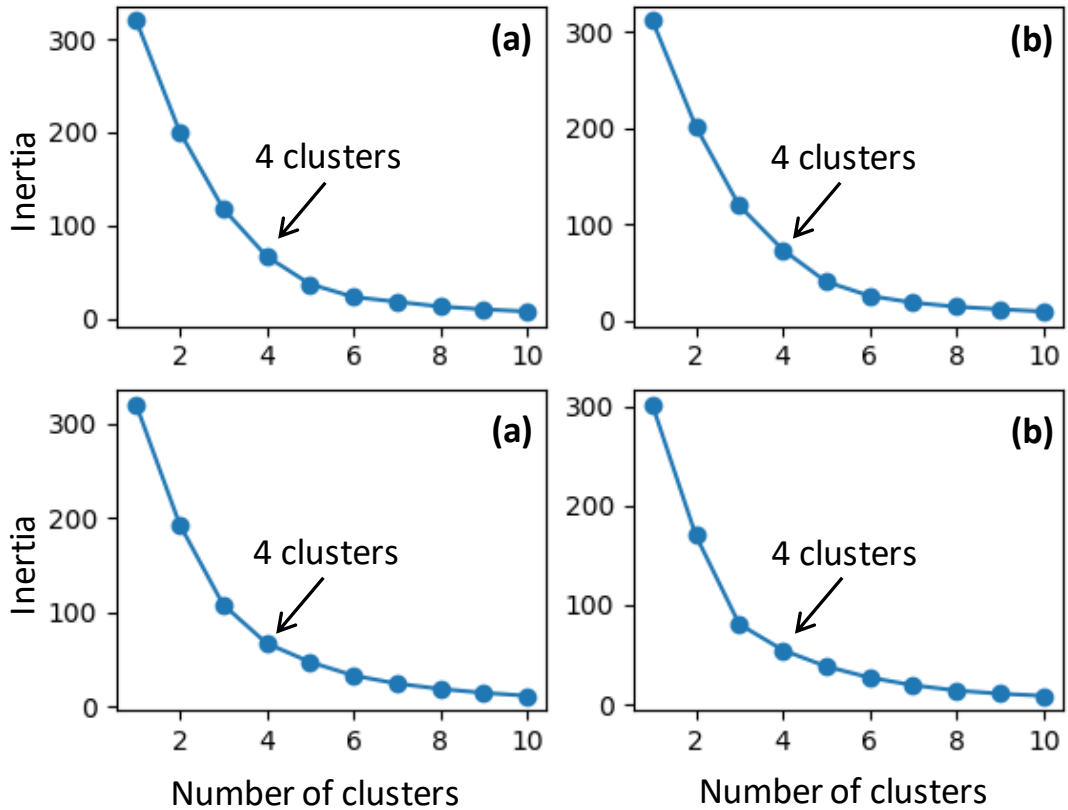

**Fig. 4** Inertia as a function of the number of clusters for data $(G, g)$ at years: (a) 2004, (b) 2009, (c) 2014, and (d) 2019. Four appears to be the optimal number.

After determining the number of clusters to be four, we cluster data points using the *K-means* method from the *sklearn library* of *Python programming*. Figure 5 depicts the clustering results for data from 2004, 2009, 2014, and 2019. We notice that almost all countries are grouped into two cluster (purple and dark green colors), while other two clusters only contain a few countries. Let us discuss further about the results.

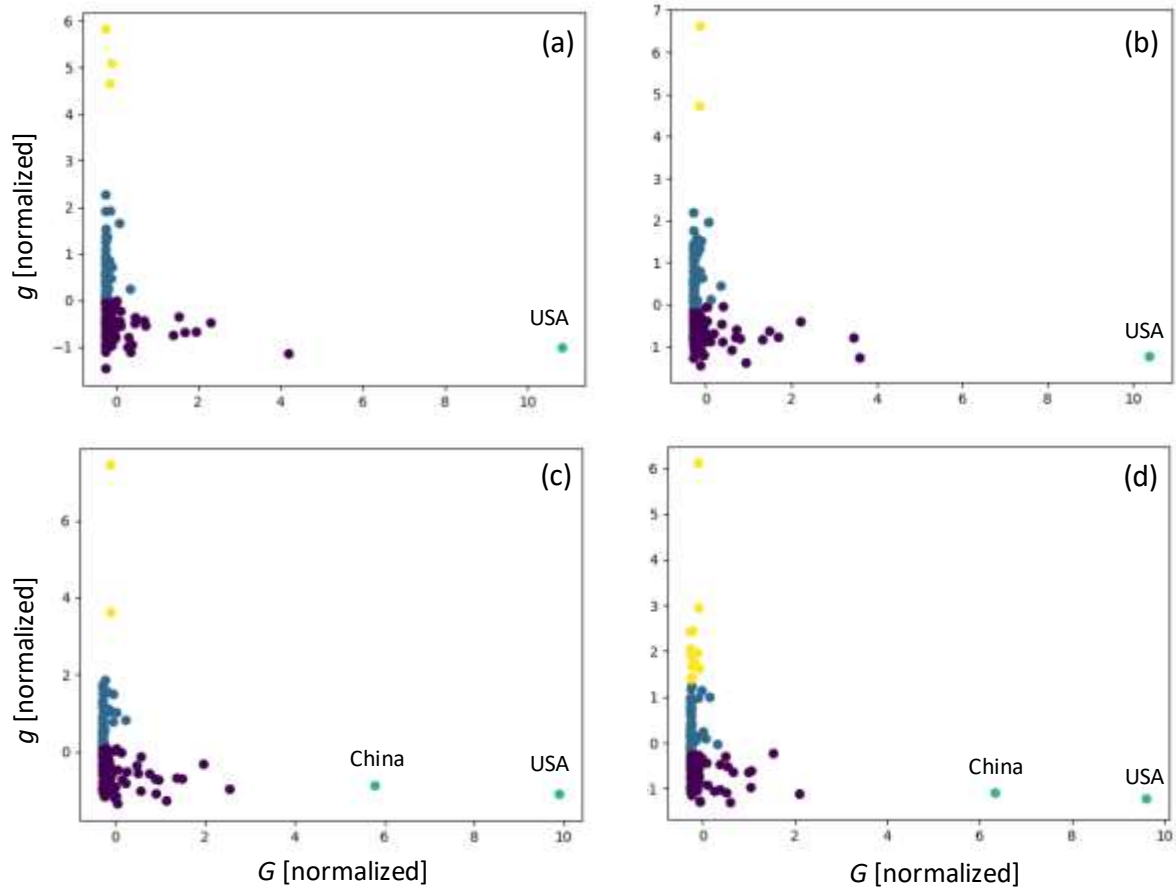

**Fig. 5** The results of clustering the points $(G, g)$ into four clusters for data at years: (a) 2004, (b) 2009, (c) 2014, and (d) 2019.

In 2004, almost all countries were grouped in the purple and dark green clusters, only three countries were in the yellow cluster (Hong Kong SAR, Singpore, and Guinea), and only one country was in the green cluster (the United States). The resulting clustering in 2009 was nearly the same as in 2004, just that the Guinea departed the yellow cluster. There is still only one country in the green cluster (the United States). The situation in 2014 is similar to that in 2009. Only Hong Kong SAR and Singapore remain in the yellow cluster. The green cluster is occupied by two countries after China joins the USA. China's rapid economic development has grouped this country with the United States. In 2019, the number of countries occupying the yellow cluster increased to 14 (Cambodia, Czech Republic, Guyana, Hong Kong SAR, Hungary, Lithuania, Namibia, North

Macedonia, Seychelles, Singapore, Slovak Republic, Slovenia, United Arab Emirates, and Vietnam), while China and the United States remain in the green cluster. All other developed countries occupy the purple clusters all year.

## 4. Conclusion

We coined the term international trade strength to describe a country's ability to generate international trade by leveraging its GDP. We have also established a theoretical foundation for the distribution function that is fulfilled by international trade strength and its rate. The distribution of these two quantities satisfies the Pareto distribution function, respectively $\propto 1/g^{1.5}$ and $\propto 1/f^{1.25}$. Using WITS and World Bank data, we obtain a Pareto distribution that satisfies $\propto 1/g^a$ and $\propto 1/f^b$ where $a \approx 1.66 - 1.82$ and $b \approx 1.16 - 1.33$, indicating a reasonably good agreement with the theory predictions. We discovered, using unsupervised machine learning, that countries can be grouped into four clusters based on GDP data and international trade strength. Finally, the derived equation for the time dependency of the international total trade that may be used to forecast the total trade in the coming years is likely acceptable.

**Declaration of competing interest**

The author declares that he has no known competing interests that could have appeared to influence the work reported in this paper.

**Data availability**

Data will be made available on request.

**Acknowledgement**

Not Available

## Appendix 1

### Exponential Distribution Function

The logarithmic of the maximum likelihood is

$$LL_{max} = n \ln \tilde{\lambda} - \tilde{\lambda} \sum_{i=1}^{n} x_i \tag{A1}$$

where

$$\tilde{\lambda} = \frac{n}{\sum_{i=1}^{n} x_i} \tag{A2}$$

### Lognormal Distribution Function

The logarithmic of the maximum likelihood is

$$LL_{max} = -\sum_{i=1}^{n} \ln x_i - n \ln \tilde{\sigma} - \frac{n}{2} \ln 2\pi - \sum_{i=1}^{n} \frac{(\ln x_i - \tilde{\mu})^2}{2\sigma^2} \tag{A3}$$

where

$$\tilde{\mu} = \frac{1}{n} \sum_{i=1}^{n} \ln x_i \tag{A4}$$

$$\tilde{\sigma}^2 = \frac{1}{n} \sum_{i=1}^{n} (\ln x_i - \tilde{\mu})^2 \tag{A5}$$

### Gamma distribution function

The logarithmic of the maximum likelihood is

$$LL_{max} = n\tilde{\alpha} \ln \tilde{\beta} - n \ln \Gamma(\tilde{\alpha}) - \tilde{\beta} \sum_{i=1}^{n} x_i + (\tilde{\alpha} - 1) \sum_{i=1}^{n} \ln x_i \tag{A6}$$

where

$$\tilde{\beta} = \frac{n\tilde{\alpha}}{\sum_{i=1}^{n} x_i} \tag{A7}$$

and $\tilde{\alpha}$ satisfies

$$n \ln \left( \frac{n\tilde{\alpha}}{\sum_{i=1}^{n} x_i} \right) + \sum_{i=1}^{n} \ln x_i - n\psi(\tilde{\alpha}) = 0 \tag{A8}$$

with

$$\psi(\tilde{\alpha}) = \frac{\frac{d\Gamma(\tilde{\alpha})}{d\tilde{\alpha}}}{\Gamma(\tilde{\alpha})} \tag{A9}$$

is the digamma function and $\Gamma(\alpha)$ is the gamma function.

**Pareto distributuion function**

The logarithmic of the maximum likelihood is

$$LL_{max} = n\tilde{\alpha} + n\tilde{\alpha}\ln\tilde{\beta} - (\tilde{\alpha}+1)\sum_{i=1}^{n}\ln x_i \tag{A10}$$

where

$$\tilde{\beta} = \min_{i} x_i \tag{A11}$$

and

$$\tilde{\alpha} = \frac{n}{\sum_{i=1}^{n}\ln x_i - n\ln\tilde{\beta}} \tag{A12}$$

**Weibull distribution function**

The logarithmic of the maximum likelihood is

$$LL_{max} = n\ln\tilde{\beta} - n\tilde{\beta}\ln\tilde{\alpha} - \frac{1}{\tilde{\alpha}^{\tilde{\beta}}}\sum_{i=1}^{n}x_i^{\tilde{\beta}} + (\tilde{\beta}-1)\sum_{i=1}^{n}\ln x_i \tag{A13}$$

where

$$\tilde{\alpha} = \left(\frac{1}{n}\sum_{i=1}^{n}x_i^{\tilde{\beta}}\right)^{1/\tilde{\beta}} \tag{A14}$$

and beta is a solution to the following equation

$$\frac{1}{\tilde{\beta}} - \frac{\sum_{i=1}^{n}x_i^{\tilde{\beta}}\ln x_i}{\sum_{i=1}^{n}x_i^{\tilde{\beta}}} + \frac{1}{n}\sum_{i=1}^{n}\ln x_i = 0 \tag{A15}$$

**Appendix 2**

**Table A1**

The results of AIC and BIC calculations using World Bank data for 2005 GDP (World Bank, 2023). The number of countries for which data is available is $n = 253$

| Distribution function | LLmax | k | AICc | BIC |
|---|---|---|---|---|
| exponential | -7335.223 | 1 | 14672.458 | 14675.979 |
| lognormal | -6798.915 | 2 | 13601.870 | 13601.830 |
| gamma | -7629.543 | 2 | 15263.126 | 15270.153 |
| Pareto ($\alpha \approx 0.135$) | **-6385.827** | **2** | **12775.695** | **12782.722** |
| Weibull | -6823.024 | 2 | 13650.088 | 13657.115 |

**Table A2**

The results of AIC and BIC calculations using World Bank data for 2010 GDP (World Bank, 2023). The number of countries for which data is available is $n = 256$

| Distribution function | LLmax | k | AICc | BIC |
|---|---|---|---|---|
| exponential | -7519.971 | 1 | 15041.954 | 15045.487 |
| lognormal | -7000.164 | 2 | 14004.367 | 14011.418 |
| gamma | -7061.379 | 2 | 14126.798 | 14133.848 |
| Pareto ($\alpha \approx 0.135$) | **-6578.231** | **2** | **13160.502** | **13167.533** |
| Weibull | -7021.105 | 2 | 14046.250 | 14053.301 |

**Table A3**

The results of AICc and BIC calculations using World Bank data for 2015 GDP (World Bank, 2023). The number of countries for which data is available is $n = 258$

| Distribution function | LLmax | k | AICc | BIC |
|---|---|---|---|---|
| exponential | -7612.666 | 1 | 15227.343 | 15230.884 |
| lognormal | -7079.894 | 2 | 14163.828 | 14170.894 |
| gamma | -7145.283 | 2 | 14294.589 | 14301.671 |
| **Pareto ($\alpha \approx 0.135$)** | **-6656.870** | **2** | **13317.779** | **13324.846** |
| Weibull | -6509.105 | 2 | 14210.027 | 14217.094 |

**Table A4**

The results of AICc and BIC calculations using World Bank data for 2020 GDP (World Bank, 2023). The number of countries for which data is available is $n = 250$

| Distribution function | LLmax | k | AICc | BIC |
|---|---|---|---|---|
| exponential | -7383.318 | 1 | 14768.647 | 14772.157 |
| lognormal | -6888.892 | 2 | 13781.824 | 13788.827 |
| gamma | -6949.362 | 2 | 13902.765 | 13909.767 |
| **Pareto ($\alpha \approx 0.135$)** | **-6479.767** | **2** | **12963.575** | **12970.577** |
| Weibull | -6910.249 | 2 | 13824.539 | 13831.542 |